\documentclass[final,usenatbib,useAMS]{mn2e}
\usepackage{multirow}
\usepackage{graphicx}
\usepackage{times}
\usepackage{float}
\usepackage{pdflscape}
\usepackage{graphicx}
\usepackage{threeparttable}
\usepackage{epstopdf}
\usepackage{amsmath}
\usepackage{amssymb}
\usepackage{xspace}
\title[Radiatively driven wind in PDS\,456]{Evidence for a radiatively driven disc-wind in PDS\,456?}
\author[Matzeu et al.]
{G. A. Matzeu,$^{1,2}$\thanks{Correspondence to: gabriele.matzeu@brera.inaf.it} 
J. N. Reeves, $^{2,3}$ V. Braito, $^{1}$ E. Nardini, $^4$ D. E. McLaughlin, $^2$  
\newauthor
A. P. Lobban, $^2$
F. Tombesi $^{5,6,7}$
and M. T. Costa $^2$ \\
$^1$INAF -- Osservatorio Astronomico di Brera, Via Bianchi 46, I-23807 Merate (LC), Italy\\
$^2$Astrophysics Group, School of Physical and Geographical Sciences, Keele University, Keele, Staffordshire ST5 5BG, UK\\
$^3$Center for Space Science and Technology, University of Maryland Baltimore County, 1000 Hilltop Circle, Baltimore, MD 21250, USA\\
$^4$INAF -- Osservatorio Astrofisico di Arcetri, Largo Enrico Fermi 5, I-50125 Firenze, Italy
\\
$^5$X-ray Astrophysics Laboratory, NASA/Goddard Space Flight Center, Greenbelt, MD, 20771, USA\\
$^6$Department of Astronomy and CRESST, University of Maryland, College Park, MD, 20742, USA
\\
$^7$Department of Physics, University of Rome ``Tor Vergata", Via della Ricerca Scientifica 1, I-00133 Rome, Italy}

		\newcommand{\pds}{PDS\,456\xspace}

		\newcommand{\fexxvabs}{Fe\,\textsc{xxv}\,He$\alpha$\xspace}
		
		\newcommand{\fexxviabs}{Fe\textsc{xxvi}\,Ly$\alpha$\xspace}

		\newcommand{\nh}{N_{\rm H}}
		\newcommand{\lognh}{\log(N_{\rm H}/\rm{cm}^{-2})}
		\newcommand{\logxi}{\log(\xi/\rm{erg\,cm\,s}^{-1})}

		\newcommand{\ledd}{L_{\rm Edd}}
		
		\newcommand{\mbh}{M_{\rm BH}}

		\newcommand{\fe}{Fe\,K\xspace}
        
		\newcommand{\ergs}{\rm erg\,s^{-1}\xspace}

		\newcommand{\rg}{R_{\rm g}}
		\newcommand{\kev}{\rm keV}
		
		\newcommand{\vw}{v_{\rm w}}
		\newcommand{\vwc}{v_{\rm w}/c}		
		\newcommand{\vwoc}{\frac{\vw}{c}}

		\newcommand{\rw}{r_{\rm w}}
		\newcommand{\fig}{Fig.\,}

		\newcommand{\suzaku}{\emph{Suzaku}\xspace} 
		\newcommand{\nustar}{\emph{NuSTAR}\xspace}		
		\newcommand{\xmm}{\emph{XMM-Newton}\xspace}

		\newcommand{\xmmnu}{\textit{XMM-Newton} \& \textit{NuSTAR}\xspace}

		\newcommand{\xstar}{\textsc{xstar}\xspace}

\begin{document}

\date{\today}

\pagerange{\pageref{firstpage}--\pageref{}} \pubyear{?}

\maketitle
\label{firstpage}


\maketitle
\begin{abstract}

\noindent We present a newly discovered correlation between the wind outflow velocity and the X-ray luminosity in the luminous ($L_{\rm bol}\sim10^{47}\ergs$) nearby ($z=0.184$) quasar \pds. All the contemporary \xmm, \nustar and \suzaku observations from 2001--2014 were revisited and we find that the centroid energy of the blueshifted \fe absorption profile increases with luminosity. This translates into a correlation between the wind outflow velocity and the hard X-ray luminosity (between 7--30\,keV) where we find that $\vwc \propto L_{7-30}^{\gamma}$ where $\gamma=0.22\pm0.04$. We also show that this is consistent with a wind that is predominately radiatively driven, possibly resulting from the high Eddington ratio of \pds.

\end{abstract}
\begin{keywords}
Subject headings: Black hole physics -- galaxies: active -- galaxies: nuclei -- quasars: individual (\pds) -- X-rays: galaxies 
\end{keywords}


\section{introduction}

Blueshifted \fe shell absorption lines, observed at $>7\,\kev$ in the rest-frame X-ray spectra of AGN, were first discovered in luminous quasars \citep[e.g.,][]{Chartas03,Reeves03,Pounds03}. The high velocities inferred, i.e. $\vw\gtrsim0.1c$, imply an association with an accretion disc-wind. The importance of these winds is supported by their frequent detection, as they are observed in the X-ray spectra of approximately $40\%$ of AGN (\citealt{Tombesi10,Gofford13}), suggesting that their geometry is characterized by a wide opening angle as recently confirmed in the luminous quasar \pds by \citet[][hereafter N15]{Nardini15}. These outflows are characterized by considerably high column densities ($N_{\rm H}\sim 10^{23}\,\rm cm^{-2}$) and a mean velocity $\left \langle \vw \right \rangle\sim0.1c$ \citep{Tombesi10} up to mildly relativistic values of $\sim0.2$--$0.4c$ in the most extreme cases (e.g., \citealt{Chartas02,Reeves09,Tombesi15}). The high velocities and high columns can result in a large amount of mechanical power, possibly exceeding the $0.5$--$5\%$ of the bolometric luminosity ($L$ hereafter), as required for significant AGN feedback (\citealt{King03,KingP03,DiMatteo05,HopkinsElvis10}). In principle the physical properties of these fast winds can provide more insight on the mechanism through which they are driven/accelerated. The possible scenarios are, \textit{radiatively driven winds} (e.g., \citealt{Proga00,Proga04}) and/or \textit{magneto-centrifugally driven winds} (MHD wind hereafter, e.g. \citealt{Ohsuga09,Kazanas12}). Most outflow studies so far mainly concentrated on radiatively driven winds  \citep[e.g.,][]{Sim10,Reeves14,Hagino15,Nomura16}. However, recently \citet{Fukumura17} successfully applied an MHD wind model to the microquasar GRO\,J1655--40 suggesting that a similar mechanism applies across the black hole mass and luminosity scales. 
\\
\indent The quasar \pds, located at redshift $z=0.184$ \citep{Torres97} is the most luminous ($L\sim10^{47}\,\ergs$; \citealt{Simpson99,Reeves00}) radio-quiet quasar in the local Universe ($z<0.3$). Since the first \xmm ($40$\,ks) observation carried out in 2001 \citep{Reeves03}, \pds has showed the presence of a persistent deep absorption trough in the \fe band above $7$\,keV. This can be associated to highly ionized Fe K-shell absorption with a corresponding outflow velocity of $v_{\rm w}\sim 0.25$--$0.3c$ (\citealt{Reeves09,Reeves14,Gofford14}; N15; \citealt{Matzeu16}, hereafter M16), which originates from an accretion disk wind.
In this letter 
we present a new correlation between the disc-wind velocity derived from the Fe K profile and the intrinsic X-ray luminosity, which we interpret in the framework of a radiatively driven wind. Values of $H_0=70\,\rm km\,s^{-1}\,Mpc^{-1}$ and $\Omega_{\Lambda_{0}}=0.73$ are assumed throughout this work and errors are quoted at the $1\sigma$ confidence level or $90\%$ where otherwise stated.


\begin{table}

\centering
\footnotesize

\begin{tabular}{cccccc}

\hline

Obs.                       &Satellite              &Start Date\,(UT)       &Exposure            &Ref.\\

&&&(ks)\\

\hline

\multirow{1}{*}{2001}          &\xmm               &2001-02-26 09:44        &$44.0$                &\multirow{1}{*}{(1)}\\


\multirow{1}{*}{2007}         &\xmm                &2007-09-12 01:07        &$179.0$               &\multirow{1}{*}{(2)}\\


\multirow{1}{*}{2007}          &\suzaku            &2007-02-24, 17:58       &$190.6$               &\multirow{1}{*}{(3)}\\


\multirow{1}{*}{2011}          &\suzaku            &2011-03-16, 15:00       &$125.6$               &\multirow{1}{*}{(4)}\\


\multirow{1}{*}{2013a}         &\suzaku            &2013-02-21, 21:22       &$182.3$               &\multirow{3}{*}{(5,6)}\\


\multirow{1}{*}{2013b}         &\suzaku            &2013-03-03, 19:43       &$164.8$               &\\


\multirow{1}{*}{2013c}         &\suzaku            &2013-03-08, 12:00       &$108.3$               &\\


\multirow{2}{*}{A}         &\xmm                  &2013-08-27, 04:41      &$95.7$              &\multirow{10}{*}{(7)}\\

							&\nustar               &2013-08-27, 03:41      &$43.8$              &\\


\multirow{2}{*}{B}         &\xmm                  &2013-09-06, 03:24      &$95.7$              &\\

							&\nustar               &2013-09-06, 02:56      &$43.0$             &\\


\multirow{2}{*}{C}         &\xmm                  &2013-09-15, 18:47      &$102.0$            &\\

							&\nustar               &2013-09-15, 17:56      &$44.0$              &\\


\multirow{2}{*}{D}         &\xmm                  &2013-09-20, 02:47      &$93.0$             &\\

							&\nustar               &2013-09-20, 03:06      &$58.6$            &\\
																																										

\multirow{2}{*}{E}         &\xmm                  &2014-02-26, 08:03      &$100.3$            &\\

							&\nustar               &2014-02-26, 08:16      &$109.5$             &\\


\hline
\end{tabular}

\caption{Summary of the \suzaku, \xmm and \nustar observations of \pds performed from 2001--2014. The net exposure for \nustar are for a single FPMA/B detector while \xmm and \suzaku net exposures are based on the EPIC-pn and individual XIS CCD detectors respectively. References: (1) \citet{Reeves03}, (2) \citet{Behar10}, (3) \citet{Reeves09}, (4) \citet{Reeves14}, (5) \citet{Gofford14}, (6) M16, (7) N15.}
\vspace{-5mm}
\label{tab:obs_table}
\end{table}


\begin{table*}

\centering
\footnotesize

\begin{tabular}{ccccccccc}

\hline

                       &               &\multicolumn{3}{c}{------------------~Gaussian~------------------}~~                                                    &\xstar                       &                      &                          \\

\\

Satellite              &Obs.      &E$_{\rm rest}$\,(1)            &$\vw/c$\,(2)                      &EW\,(3)                       &$\vw/c$\,(4)           &$L_{2-10\,\kev}$\,(5)    &$L_{7-30\,\kev}$\,(6)       \\

                       &                &(keV)                     &                                &(eV)                     &                             &$(\times10^{44}\,\ergs)$                        &$(\times10^{44}\,\ergs)$                \\
 
\hline

\\

\multirow{5}{*}{\xmmnu}         
  
                    &A           &$9.25_{-0.11}^{+0.18}$     &$0.286_{-0.010}^{+0.017}$        &$-88\pm47$              &$0.299_{-0.012}^{+0.012}$    &$7.94\pm0.26$  &$5.22\pm0.11$\\

			         &B           &$9.06_{-0.20}^{+0.19}$     &$0.268_{-0.020}^{+0.019}$        &$-295\pm63$             &$0.278_{-0.020}^{+0.016}$    &$3.57\pm0.31$  &$1.68\pm0.06$\\

                    &C           &$9.12_{-0.09}^{+0.11}$     &$0.275_{-0.009}^{+0.011}$        &$-359\pm55$             &$0.286_{-0.007}^{+0.006}$    &$4.51\pm0.26$  &$2.60\pm0.08$\\

			         &D           &$9.16_{-0.12}^{+0.12}$     &$0.277_{-0.012}^{+0.012}$        &$-256\pm60$             &$0.292_{-0.007}^{+0.008}$    &$5.10\pm0.31$  &$2.93\pm0.08$\\

			         &E           &$8.91_{-0.11}^{+0.11}$     &$0.268_{-0.020}^{+0.019}$        &$-405\pm60$             &$0.269_{-0.009}^{+0.008}$    &$3.68\pm0.35$  &$1.81\pm0.06$\\

\\
			         
\multirow{5}{*}{\suzaku}

			         &2007       &$9.26_{-0.13}^{+0.15}$     &$0.287_{-0.013}^{+0.015}$        &$-235\pm74$             &$0.296_{-0.012}^{+0.012}$    &$4.35\pm0.34$  &$2.32\pm0.12$          \\

			         &2011       &$9.07_{-0.20}^{+0.18}$     &$0.270_{-0.020}^{+0.018}$        &$-460\pm151$            &$0.266_{-0.011}^{+0.011}$    &$4.14\pm0.53$  &$2.21\pm0.19$          \\

			         &2013a      &$8.85_{-0.29}^{+0.33}$     &$0.248_{-0.030}^{+0.035}$        &$>-214$                 &$0.248_{-0.020}^{+0.030}$    &$3.53\pm0.35$  &$1.88\pm0.14$          \\

			         &2013b      &$8.86_{-0.09}^{+0.09}$     &$0.249_{-0.010}^{+0.010}$        &$-390\pm81$             &$0.255_{-0.010}^{+0.010}$    &$2.90\pm0.36$  &$1.55\pm0.16$          \\

			         &2013c      &$8.70_{-0.08}^{+0.07}$     &$0.234_{-0.009}^{+0.008}$        &$-568\pm82$             &$0.240_{-0.010}^{+0.010}$    &$2.77\pm0.38$  &$1.46\pm0.17$          \\

\\

\multirow{2}{*}{\xmm}

			         &2001       &$9.84_{-0.16}^{+0.19}$     &$0.335_{-0.019}^{+0.018}$        &$-337\pm94$             &$0.340_{-0.015}^{+0.018}$    &$10.5\pm0.07$           &$4.97\pm0.28$          \\

                    &2007       &$9.51_{-0.14}^{+0.14}$     &$0.309_{-0.013}^{+0.013}$         &$>-71$                  &$0.299_{-0.034}^{+0.026}$    &$5.55\pm0.29$           &$3.67\pm0.14$          \\

\\

\hline
\end{tabular}

\caption{Summary of the observations and the corresponding properties of the \fe absorption features. These are obtained by fitting with a Gaussian profile (columns\,1--3) and then subsequently with \xstar (columns\,4--6). From the former we measured with the respective $90\%$ errors on: (1)\,the rest-frame energy, (2)\,the inferred outflow velocity and (3)\,the equivalent width of each line. By modelling with \xstar we measured the outflow velocity\,(4), while from the fits, the column density ranges from $\lognh=23.16_{-0.23}^{+0.28}$--$24.06_{-0.21}^{+0.18}$ and ionization ranges from $\logxi=5.1_{-0.21}^{+0.11}$--$6.33_{-0.19}^{+0.08}$. In columns\,(5) and (6) we report the intrinsic luminosities and their corresponding $1\sigma$ errors in the $2$--$10\,\kev$ and $7$--$30\,\kev$ bands, respectively.}

\label{tab:xmm/nustar}
\end{table*}

\section{Wind Analysis}

\begin{figure}
\begin{center}
\includegraphics[width=0.38\textwidth]{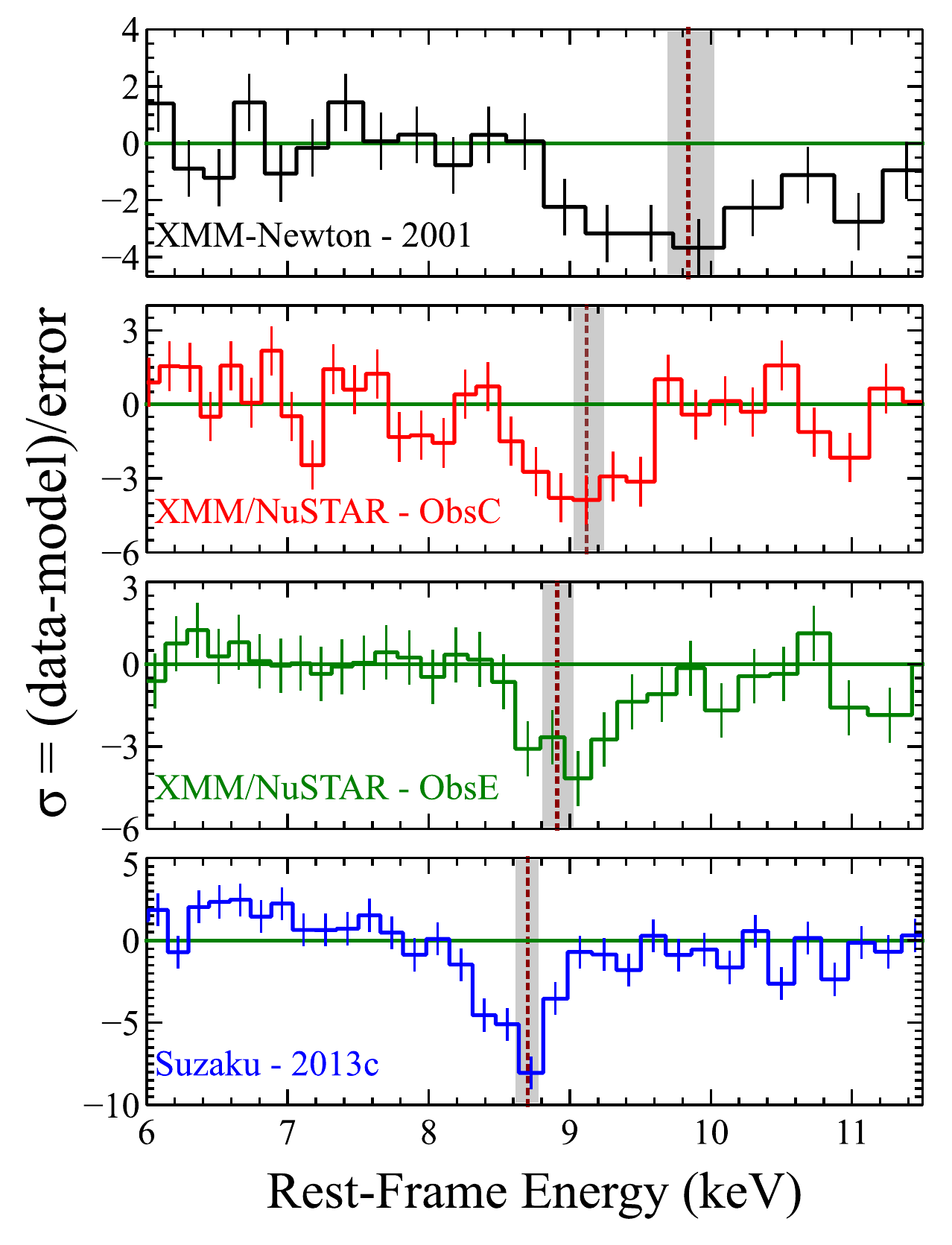}
\caption{Residuals to the power-law continuum fitted between $2$--$10\,\kev$ band from four selected observations of \pds decreasing in X-ray luminosity. This shows how the centroid energy of the \fe absorption profile generally decreases in energy with decreasing luminosity from top to bottom. The vertical dotted lines and the gray shadowed areas indicate the position of the centroid and the $90\%$  uncertainty (see Table\,\ref{tab:xmm/nustar}).}
\label{fig:pds_letter_fek_del}
\end{center}
\end{figure}

We utilize all the available \xmm, \suzaku and \nustar datasets, concentrating on the X-ray band above $2\,\kev$ fitted with a simple power-law continuum modified by Galactic absorption ($\nh=2.0\times10^{21}\,\rm cm^{-2}$) and a single layer of neutral partial covering absorption to account for any spectral curvature with velocity set to be equal to the fast wind's outflow velocity (see M16). A broad \fe emission line at $7\,\kev$ was also included, where the width was fixed to that measured by the absorption line when the latter is modelled with a simple Gaussian. The X-ray spectra were adopted from previous work (see Table\,\ref{tab:obs_table} for details). In \fig\ref{fig:pds_letter_fek_del}, the residuals of the \fe absorption profiles as a ratio to a power-law continuum are plotted for four of the twelve observations with different X-ray luminosities (in decreasing order). It is evident that the centroid energy increases across the different epochs with increasing luminosity.

As an initial test we investigate how the energy of the \fe profile responds to the ionizing luminosity, measured in either the 7--30\,keV band ($L_{7-30}$ hereafter) or the 2--10\,keV band ($L_{2-10}$) between all the sequences. The $7$--$30\,\kev$ band was chosen as it is above the \fe absorption edge threshold and is where the primary continuum emerges. This luminosity is absorption corrected and is either inferred directly from the \nustar data or from extrapolating the best fit model above $10\,\kev$, for those sequences where no \nustar observation is available\footnote{The extrapolation can be considered a good assumption, given that all the \nustar observations of \pds indicate a continuum shape of a simple power-law ($\Gamma=2.2$--$2.6$) without any strong hard excess.}. 
Monte Carlo simulations were performed in order to estimate the uncertainties on the absorption-corrected intrinsic luminosity. For each observation, 1000 simulations were performed with the \textsc{xspec} command \texttt{fakeit}, assuming the best-fit models obtained from the spectral analysis of the actual data. Each simulated spectrum was fitted again to take into account the uncertainties on the null hypothesis, allowing the main spectral parameters vary. These are the photon index and normalization of the continuum, the covering fraction and column density of the partial covering absorber, as well as the column and outflow velocity of the ionized absorber. With the new model, a second simulation was run and fitted, from which we derived the distribution of the luminosities, which were well approximated by a Gaussian distribution. The subsequent $1\sigma$ uncertainty was used as the input to the correlation analysis. We also extracted the simulated distributions of the main spectral parameters to check that they were representative of the uncertainties of the actual best fit models. Thus the derived errors on the luminosities fully account for the uncertainties on each of the free spectral parameters.

\begin{figure}
\begin{center}
\includegraphics[width=0.39\textwidth]{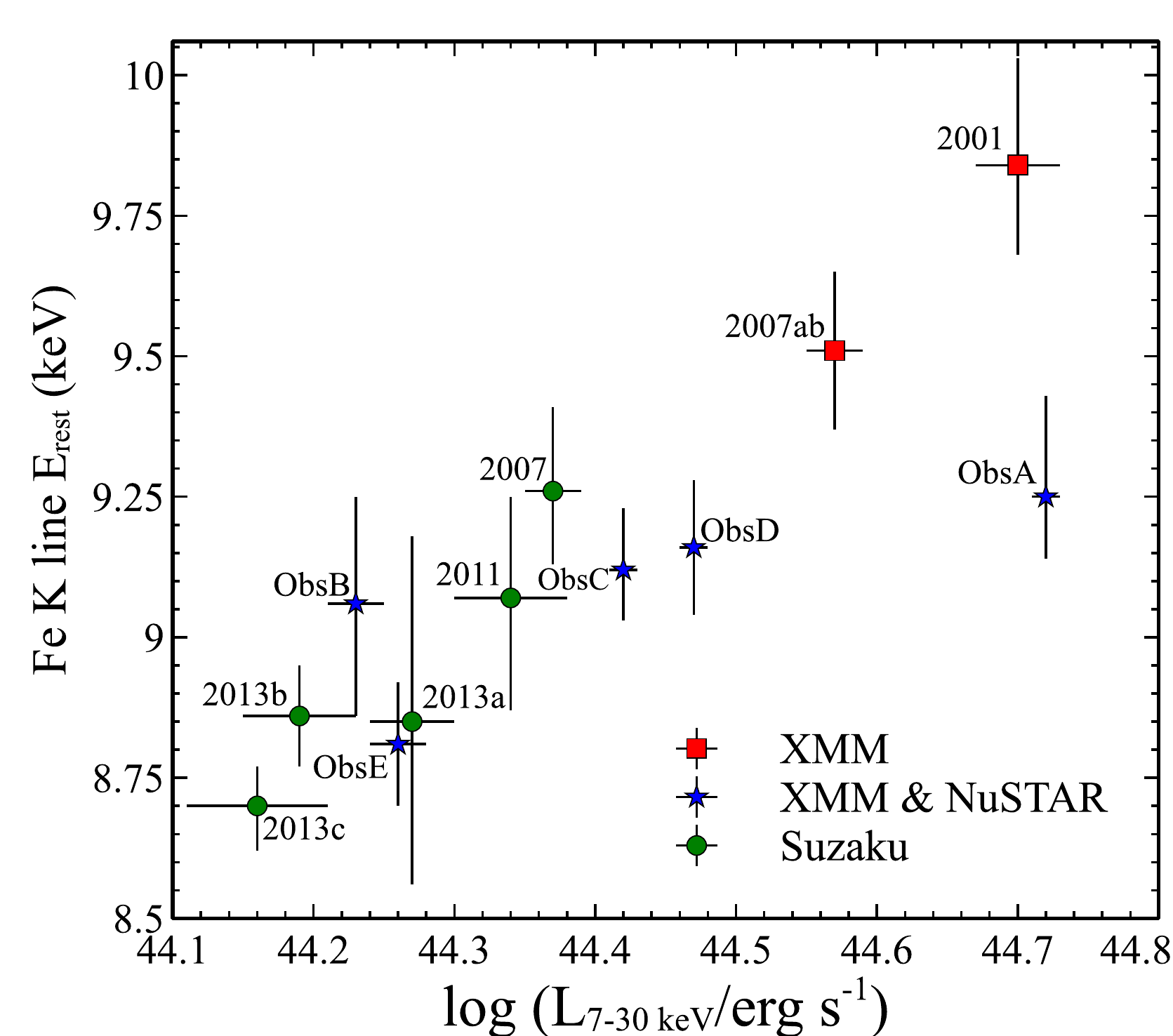}
\caption{Plot showing how the centroid energy of the \fe features correlates with $L_{7-30}$ in \pds. The distribution of the data favours a positive linear correlation between the two quantities. \suzaku (green circles), \xmmnu (blue stars) and \xmm (red squares).}
\label{fig:thesis_feK_E_logL}
\end{center}
\end{figure}

\begin{figure}
\begin{center}
\includegraphics[width=0.35\textwidth]{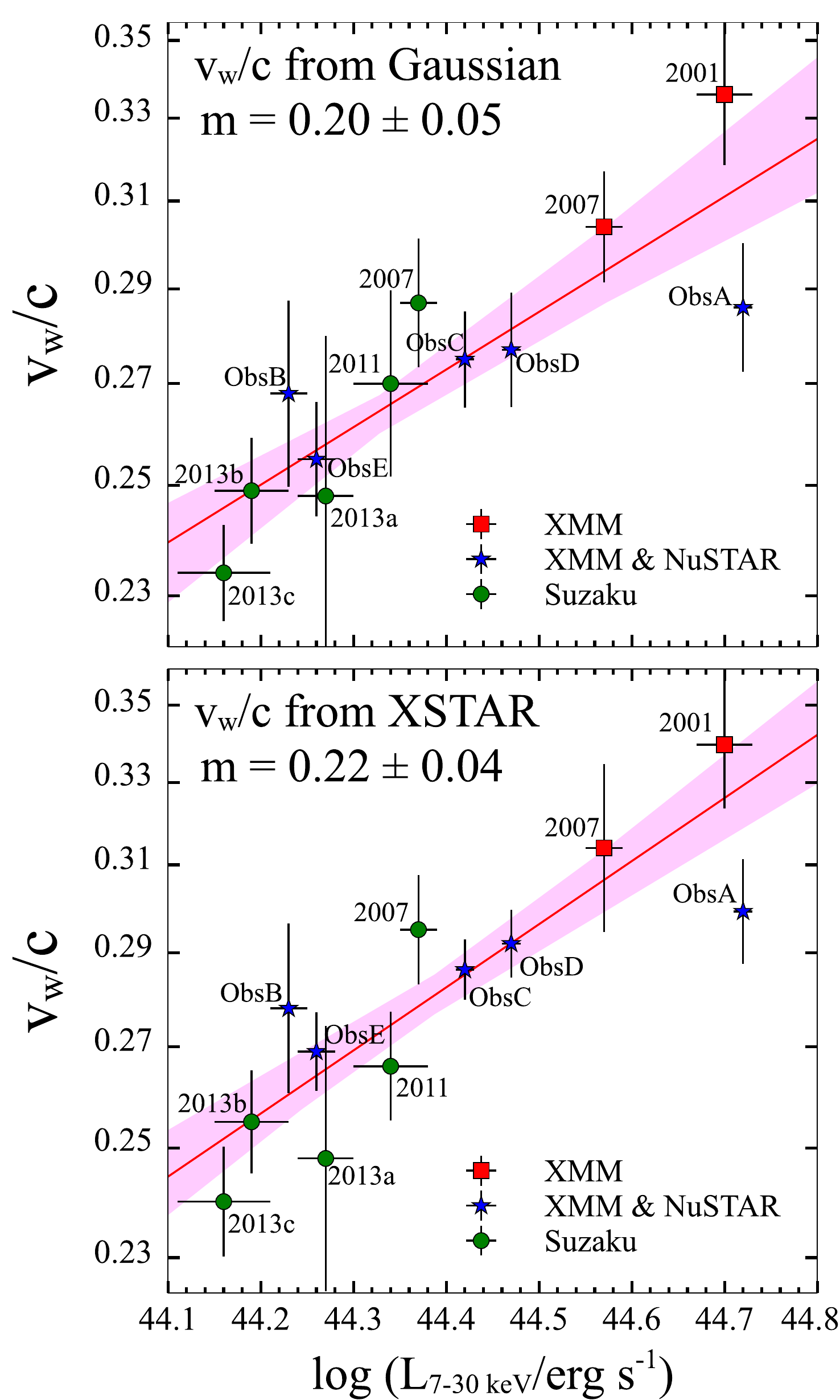}
\caption{Top: Log--log plot showing the correlation between the outflow velocity, derived from fitting the \fe features with a Gaussian profile and the intrinsic $7$--$30\,\kev$ continuum luminosity ($L_{7-30}$). The red line shows the linear regression fitted to the data with $1\sigma$ uncertainty shaded. Bottom: the outflow velocity is now measured by fitting the \fe profile with an \xstar grid where the linear regression produced a gradient of $0.22\pm0.04$. These positive correlations suggest that when the source is more luminous the disc-wind is faster.}
\label{fig:thesis_feK_logvw_logL}
\end{center}
\end{figure}

We first parametrize the blueshifted absorption lines with a Gaussian profile, whose main parameters are tabulated in Table\,\ref{tab:xmm/nustar}. In \fig\ref{fig:thesis_feK_E_logL} we show the centroid energy versus $L_{7-30}$ which becomes more blueshifted at higher luminosities suggesting a positive correlation with the intrinsic luminosity. 
Thus from the centroid energy we can infer the corresponding outflow velocity ($\vwc$) of the material, assuming it is mainly associated with \fexxviabs rather than \fexxvabs (see Table\,\ref{tab:xmm/nustar}). The top panel of \fig\ref{fig:thesis_feK_logvw_logL} shows the outflow velocity derived from the Gaussian fit against $L_{7-30}$ plotted in log--log space. Subsequently the data points were fitted with a standard regression line of the form $\log(\vwc)=m\log(L_{7-30})+\log(C)$, performed through the bivariate correlated errors and intrinsic scatter algorithm (\textsc{bces}, \citealt{AkritasBershady96}) which is one of the most common methods that takes into account errors in both x and y values. The regression line produced a gradient of $m=0.20\pm0.05$ where the shaded magenta area indicates the slope dispersion (at $1\sigma$ confidence level). A comparable regression fit is found with the intrinsic 2--10\,keV luminosity ($L_{2-10}$), with $m=0.22\pm0.04$ (see Table\,\ref{tab:regressions}). 
\\
\indent The \fe absorption features were then modelled with a more physically motivated self-consistent \textsc{xstar} grid, previously adopted in N15 and M16, generated using the UV to hard X-ray SED of \pds. The \xstar model allows the column and/or the ionization to adjust between observations, as well as the net outflow velocity. The corresponding $\vwc$ values of all twelve observations derived from the \xstar model are listed in Table\,\ref{tab:xmm/nustar} and plotted against $L_{7-30}$ on the bottom panel in Fig.\,\ref{fig:thesis_feK_logvw_logL}. The linear regression fit gives a gradient of $m=0.22\pm0.04$ consistent with the Gaussian analysis. Similarly, the correlation with $L_{2-10}$ also gives a consistent result of $m=0.24\pm0.03$. From both the correlations shown in Fig.\,\ref{fig:thesis_feK_logvw_logL} it is clear that the outflow velocity of the highly ionized material is indeed responding to the intrinsic luminosity, where the most luminous epochs of \pds are characterized by a faster outflow (e.g., $\vwc=0.296\pm0.012$ in the 2007 \suzaku observation and $\vwc=0.340_{-0.015}^{+0.018}$ in the 2001 \xmm observation). On the other hand, the low-luminosity 2013 \suzaku observations (i.e. 2013b or 2013c, see Tables\,\ref{tab:obs_table} and \ref{tab:xmm/nustar}) present overall the slowest outflows with $\vwc=0.255\pm0.010$ and $\vwc=0.240\pm0.010$, respectively. Thus an important question is: how the changes in luminosity have a direct impact on the outflow velocity in the disc-wind observed in \pds?

\begin{table}

\centering
\footnotesize

\begin{tabular}{cccccc}

\hline

$x$                       &$y$                      &$m$                 &$\Delta m$              &$\chi_{\nu}^2$             &$P_{\rm null}$ \\

\hline

\multicolumn{6}{c}{From Gaussian}\\

$\log(L_{2-10})$         &\multirow{2}{*}{$\log(\vw/c)$}          &$0.22$             &$0.04$                   &$0.84$                    &\multirow{2}{*}{$6.70\times10^{-7}$}\\


$\log(L_{7-30})$		 &                                         &$0.20$             &$0.05$                   &$1.08$                   \\

\\

\multicolumn{6}{c}{From \xstar}\\

$\log(L_{2-10})$	    &\multirow{2}{*}{$\log(\vw/c)$}          &$0.24$              &$0.03$                  &$0.82$                     &\multirow{2}{*}{$8.29\times10^{-8}$}\\

$\log(L_{7-30})$	    &                                        &$0.22$              &$0.04$                  &$1.01$                      \\

\\
\hline

\end{tabular}

\caption{Results of the linear regressions between the measured outflow velocities (measured from both Gaussian profile and \xstar grids) and the intrinsic luminosities. $m$ and $\Delta m$ are the gradient and the respective $1\sigma$ errors respectively, whereas $\chi_{\nu}^{2}$ is the reduced $\chi^2/{\rm dof}$. $P_{\rm null}$ is the null hypothesis probability that there is no linear correlation.
}
\vspace{-5mm}

\label{tab:regressions}
\end{table}

\section{Discussion}

\subsection{A radiatively driven wind?}
\label{sub:Can we account for the variability in the wind velocity?}

The positive correlation between the outflow velocity and the ionizing luminosity of \pds, may imply that the wind is radiatively driven. The high ionization derived from the wind seen at \fe in \pds, with ionization parameters typically of $\logxi\sim5$--$6$ (e.g., N15), might result in a small line opacity, due to only Fe\,\textsc{xxv-xxvi}. 
While at first sight this suggests that the dominant interaction mechanism within the photon field is through Thompson scattering (e.g., \citealt{KingP03}), the opacity may be further boosted by the presence of lower ionization X-ray absorption within the wind. This lower ionization gas towards \pds has been observed in the form of the clumpy X-ray partial covering absorber with size scales of $\sim10s$ of $\rg$ from its variability (M16), as well as through the presence of broad absorption profiles revealed 
in the soft X-ray band \citep{Reeves16}.
Furthermore blueshifted UV emission and absorption lines from the wind are also present in \pds \citep{O'Brien05}, similar to those in BAL quasars. Such gas would boost the radiation force acting on the wind, via line driving \citep[see][]{Hagino16a}, resulting in an increase in the force multiplier factor.

To quantify further, we start from the definition of the net force (i.e., radiative minus gravitational force) acting on an electron--ion pair within the wind:

\begin{equation}
F_{\rm net}=\mu m_{\rm p} v(\rw) \frac{\mathrm{d} v}{\mathrm{d} \rw}= \frac{L\,\sigma}{4\pi \rw^{2} c} - \frac{G\mbh \mu m_{\rm p}}{\rw^{2}},
\label{eq:netforce}
\end{equation}
where $\mu=n_{\rm H}/n_{\rm e}\sim1.2$ is a constant factor accounting for cosmic elemental abundances, $m_{\rm p}$ is the proton mass, $r_{\rm w}$ is the radial distance from the central X-ray source and $\sigma=\sigma_{\rm PE} + \sigma_{\rm T}$ is the total cross-section (photoelectric plus Thomson). Rearranging equation\,(\ref{eq:netforce}), gives the equation of motion within the wind:

\begin{equation}
v(\rw)\mathrm{d} v = \left( \frac{L\,\sigma}{4\pi \rw^{2}\mu m_{\rm p} c} - \frac{G\mbh}{\rw^{2}}\right) \mathrm{d} \rw.
\label{eq:eqparticle_in_wind}
\end{equation}
Now we define $\rw=R_{\rm in}$ as the launch radius and $v=v_{i}$ as the initial velocity of the disc-wind. At larger radii we assume that $\rw \gg R_{\rm in}$ and hence $\rw \to \infty$ and $v \to \vw$ (i.e., the terminal speed at large radii). Integrating equation\,(\ref{eq:eqparticle_in_wind}) between these limits and rearranging gives:-

\begin{equation}
\vw^{2} - v_{i}^{2} = \frac{2G\mbh}{R_{\rm in}}\left( \frac{L\,\sigma}{4\pi G\mbh \mu m_{\rm p}c} - 1 \right).
\label{eq:integrated_eqparticle}
\end{equation}
On the right hand side, the first term is the escape velocity of the particle and the second term is essentially the local $L/\ledd$ within the wind, where the condition $\sigma > \sigma_{\rm T}$ results from the possible contribution of line driving within the wind, giving a force multiplier factor $>1$. Thus if we simplify the above equation such that $\vw^{2} \gg v_{i}^{2}$ and that locally $L > \ledd$  (e.g., as a result of the force multiplier), we get a dependence between the outflow velocity, the luminosity and $R_{\rm in}$:

\begin{equation}
\vwoc \sim \left ( \frac{\sigma}{2\pi \mu m_{\rm p}c^3} \right )^{1/2} k_{7-30}^{1/2} L_{7-30}^{1/2} R_{\rm in}^{-1/2} \propto k_{7-30}^{1/2} L_{7-30}^{1/2}R_{\rm in}^{-1/2}.
\label{eq:vw_radiative}
\end{equation}
Here $L = k_{7-30} L_{7-30}$, where $k_{7-30}$ is the bolometric correction factor ($\gtrsim100$ for \pds).

Our relationship expressed in equation\,(\ref{eq:vw_radiative}) between the outflow velocity and the ionizing luminosity is `steeper' than what is observed in \fig\ref{fig:thesis_feK_logvw_logL}. There are two possibilities that may explain this; firstly the wind launch radius, $R_{\rm in}$, may not be constant with luminosity, or $k_{7-30}$ is variable. In the first scenario as the luminosity increases, the innermost parts of the wind can become fully ionized. Subsequently at higher luminosities the innermost fastest streamlines of the wind become effectively unobservable. Given that the ionization parameter varies as $\xi \propto L\,r_{\rm w}^{-2}$, the effective radius at which the wind is observable increases with luminosity, leading to the `flattening' of the slope as observed. 

In the latter case the terminal velocity should be proportional to the whole bolometric luminosity rather than just what is directly measured in the hard X-ray band. Furthermore, all of the \xmm observations of \pds clearly show that the UV flux is considerably less variable than in the X-ray band \citep[see Fig. 2][]{Matzeu17}. As a result $k_{7-30}$ might be variable as a function of luminosity which again may `flatten' the expected correlation. While it is not currently possible to quantitatively test the relation between $\vwc$ and $L$ (as only the \xmm observations have simultaneous optical/UV photometry), the resulting correlation between $\vwc$ and $L$ is at least qualitatively consistent with a steeper slope of $\sim0.5$. Ongoing optical/UV and X-ray monitoring of \pds with \textit{Swift} will allow us to measure the bolometric correction as a function of luminosity

One final question is whether it is plausible to radiatively accelerate the wind up to $0.3c$? From equation\,(\ref{eq:integrated_eqparticle}) and taking $L\sim L_{\rm Edd}$ for \pds, then for a reasonable launch radius of $30\,\rg$ (given the rapid wind variability, (M16), then only a modest force multiplier factor of  $\tau=\sigma/\sigma_{\rm T}=2.5$ is required. Furthermore, if \pds is mildly super-Eddington (e.g. $2L_{\rm Edd}$), then $\tau$ approaches unity. Thus the observed velocities can be easily reproduced within this framework. Note that a comparable positive correlation between the outflow velocity and luminosity has been recently found by \citet{Gofford15} and \citet{Fiore17}, the latter from collating a sample of known AGN outflows.


\subsection{MHD launching scenario}

Alternatively, could MHD winds reproduce the above correlation? While to date there is no expected correlation in the MHD wind scenarios (e.g., \citealt{Ohsuga09,Kazanas12}), the observed velocity can still depend on the continuum photon index, the SED shape and in particular the UV to X-ray slope, $\alpha_{\rm ox}$. 

Indeed \citet{Fukumura10} predicted that sources with steeper $\alpha_{\rm ox}$ values (i.e. X-ray quiet compared to the UV) should be characterized by faster disc-winds. This is the result of the shape of the ionizing SED. When the X-ray luminosity increases compared to the UV, the innermost parts of the wind become fully ionized. As a result only the slower streamlines of the wind are observed, launched from further out. In this scenario, we may expect the outflow velocity to decrease with increasing X-ray luminosity, the opposite of what is observed in \pds.

Nonetheless the strong positive correlation in Fig.\,\ref{fig:thesis_feK_logvw_logL} does not preclude that MHD processes play an important role in providing the initial lift of the wind material off the disc (i.e., $v_i$). In this regard we have some evidence that confirms this scenario, where in the low-luminosity observation of \pds in 2013 with \suzaku, a strong X-ray flare increased the luminosity by a factor of $\sim4$ in just $\sim50\,\rm ks$ which was likely to be magnetically driven (M16). The disc-wind became stronger after the flare was observed, supporting a scenario whereby MHD processes contributed at least to the initial ejection of the material off the disc, only for it to be then accelerated by the intense radiation pressure. Other AGN that accrete near the Eddington limit, e.g. in luminous quasars and Narrow Line Seyfert\,1 Galaxies such as: APM\,08279$+$5255 \citep{Chartas03}, PG1211$+$143 \citep{Pounds03},  IRAS\,F11119$+$3257 \citep{Tombesi15}, 1H\,0707--495 \citep{Hagino16a}, IRAS\,13224--3809 \citep{Parker17}, are also likely to be promising candidates for radiatively driven winds. However the contribution of MHD winds are likely to be more important at the lower $L/\ledd$ ratio end of the population, where radiation driving would be difficult without large force multiplier factors.

\section{Conclusions}

We analysed all twelve X-ray observations of the fast disc-wind in \pds from 2001 to 2014, observed  through its iron K absorption profile. For the first time we find a positive correlation between the wind velocity and the intrinsic hard X-ray luminosity. 
This provides evidence that the accretion disc-wind in \pds is most likely to be radiatively driven, with higher luminosities helping to accelerate the wind to higher terminal velocities.

\section{acknowledgements}

We thank the anonymous referee for their helpful report. GM and VB acknowledge support from the Italian Space Agency (ASI INAF NuSTAR I/037/12/0). JR, DML and AL acknowledge the support of STFC. EN is supported by EU's Marie Sk{\l}odowska-Curie grant no. 664931.  FT acknowledges support by the Programma per Giovani Ricercatori - 2014 ``Rita Levi Montalcini''.

\bibliographystyle{mn2e}
\bibliography{matzeu_references}

\end{document}